\documentstyle[preprint,aps]{revtex}
\begin{document}
\tighten
\draft
\preprint{PSU/TH/193; hep-ph/9801384}
\title {SECOND ORDER IN MASS RATIO RADIATIVE-RECOIL CORRECTIONS 
TO HYPERFINE SPLITTING IN MUONIUM} 
\author {Michael I. Eides \thanks{E-mail address:  
eides@phys.psu.edu, eides@lnpi.spb.su}}
\address{ Department of Physics, Pennsylvania 
State University, 
University Park, PA 16802, USA\thanks{Temporary address.}\\ 
and
Petersburg Nuclear Physics Institute,
Gatchina, St.Petersburg 188350, Russia\thanks{Permanent address.}}
\author{Howard Grotch\thanks{E-mail address: h1g@psu.edu}}
\address{Department of Physics, Pennsylvania State University,
University Park, PA 16802, USA}
\author{Valery A. Shelyuto \thanks{E-mail address: 
shelyuto@onti.vniim.spb.su}} 
\address{D. I.  Mendeleev Institute of Metrology, 
St.Petersburg 198005, Russia}
\date{January, 1998}

\maketitle
\begin{abstract}
Radiative-recoil corrections to hyperfine splitting in muonium of orders 
$\alpha(Z\alpha)(m/M)^2E_F$ and $(Z^2\alpha)(Z\alpha)(m/M)^2E_F$ are 
calculated. These corrections are of the second order in the small 
electron-muon mass ratio. An analytic expression $[(-6 \ln2- 
\frac{3}{4})\alpha (Z\alpha) - \frac{17}{12} (Z^2 
\alpha) (Z\alpha)](\frac{m}{M})^2 E_F$ is obtained. 
Numerically the correction is equal to  $-0.0351\:\mbox{kHz}$ and is of 
the same order of magnitude as the expected accuracy of the 
current Los Alamos experiment to measure the hyperfine splitting.  
\end{abstract} 


\section{Introduction}
\label{intro}

The hyperfine splitting in the ground state of muonium has been measured 
with high precision \cite{mbb}

\begin{equation}
\Delta\nu_{\text{Mu}}(n=1) = 4~463~302.88~(16)~{\rm kHz}\qquad 
\delta=3.6\cdot 10^{-8}, 
\end{equation} 

and there are further opportunities for improving the accuracy. The lifetime 
of the higher hyperfine state is extremely large 
$\tau=1\cdot 10^{13}~{\rm s}$ and gives negligible contribution to the 
linewidth.  Thus the linewidth is completely defined by the muon lifetime 
$\tau_\mu\approx 2.2\cdot 10^{-6}~{\rm s}$ which leads to the width 
$\Gamma_\mu/h=72.3~{\rm kHz}$. The experimental accuracy attained thus far 
corresponds to measuring the energy splitting with an experimental error of 
$\Delta\nu_{exp}/(\Gamma_\mu/h)\approx 2.2\cdot 10^{-3}$ of the natural 
linewidth. A new experiment is now running at Los Alamos \cite{vh}, 
with the goal of measuring $\Delta\nu_{\text{Mu}}(n=1)$ with a precision of 
about $1\cdot 10^{-8}$, or at the level of a few hundredths of a kilohertz.

In anticipation of the long-awaited results from this precision experiment, 
recent years have witnessed a surge of theoretical activities in this field. 
A tedious calculation of the purely radiative corrections of order 
$\alpha^2(Z\alpha)E_F$ 
\cite{eks1,eks2,eks3,eks4,kn1,es,kn2}  
was followed by works  \cite{kar1,kn1,pl,bcs} where leading corrections of 
higher orders ($\alpha(Z\alpha)^3\ln Z\alpha\; E_F$, 
$\alpha^2(Z\alpha)^2\ln^2 Z\alpha\; E_F$, $\alpha^2(Z\alpha)^2\ln Z\alpha\;  
E_F$, $(Z\alpha)^3(m/M)\ln^2 Z\alpha\; E_F$, $(Z\alpha)^3(m/M)\ln 
Z\alpha\ln(m/M)E_F$, \\$\alpha(Z\alpha)^2(m/M)\ln^2Z\alpha\;E_F$) were 
obtained. Some other corrections of higher orders 
($\alpha^2(Z\alpha)(m/M)\ln^3(m/M)\;E_F$, 
$\alpha^2(Z\alpha)(m/M)\ln^2(m/M)\;E_F$) have already been known for some 
time \cite{es0,eksr3}. The magnitude of these higher order corrections 
varies from a few thousandths to a few tenths of a kilohertz and some of 
them turned out to be rather large. The correction of order 
$\alpha(Z\alpha)^2E_F$ was also recalculated recently \cite{pach,kn3}. 

In this paper we consider radiative-recoil corrections of order 
$\alpha(Z\alpha)(m/M)^2E_F$ and $(Z^2\alpha)(Z\alpha)(m/M)^2E_F$, which were 
not discussed so far. For muonium both the fine structure constant 
$\alpha$ and the electron-muon mass ratio $(m/M)$ are of the same order of 
magnitude, and one could anticipate that the corrections under discussion 
would produce contributions comparable to the higher order corrections of 
order $\alpha^4\;E_F$ mentioned above. Another reason for interest in the 
radiative-recoil correction of second order in mass ratio is connected with 
a long standing discrepancy between the results of \cite{sty} and 
\cite{eks5,beks} for radiative-recoil corrections of the first 
order in mass ratio, induced by the radiative insertions in the electron 
line. The discrepancy is about three standard deviations of the accuracy of 
the numerical calculations in \cite{sty}, and is equal to $0.22$ kHz. This 
discrepancy, which was of a purely academic interest when the 
respective calculations were performed, now acquires some practical 
importance. An apparent reason for this contradiction could be connected 
with the fact that while in \cite{eks5,beks} an explicit 
expansion in the small mass ratio was performed, with only the contribution 
of the first order preserved, in the calculations in \cite{sty} only the 
nonrecoil contribution was thrown away, and consequently the result of this 
work contains terms linear in the mass ratio as well as all 
contributions of higher orders in the mass ratio. With the results of this 
work we will be able to test this hypothesis about the origin of the 
discrepancy between the results in \cite{sty} and \cite{eks5,beks}.

Radiative-recoil corrections to hyperfine splitting are generated by the 
diagrams with all possible radiative insertions in the two-photon exchange 
graphs. It suffices to calculate respective matrix elements 
between the Coulomb-Schr\"odinger wave functions (see, e.g., discussion in
\cite{eksann1}). We will consider below separately the corrections generated 
by the three types of diagrams: polarization insertions in the exchanged 
photons, radiative insertions in the electron line, and radiative insertions 
in the muon line.

\section{Polarization Operator Contributions}                                               
\label{polarsect}

The general expression for the $\alpha(Z\alpha)E_F$ corrections to hyperfine 
splitting induced by the electron vacuum polarization insertions in the 
exchanged photons has the form (see, e.g., \cite{eksann1})

\begin{equation}         \label{polgen}
\delta E^{\text{pol}}~=~ \alpha (Z\alpha) E_F~  \frac{1}{\pi^2 \mu }
\int_0^1 {dv}\int \frac{d^4 k}{i \pi^2}~\frac{1}{k^2} 
\frac{v^2 (1 - v^2/3)}{4 - k^2 (1 - v^2)} 
\end{equation}
\[
\left(\frac{1}{k^2 + \mu^{-1}k_0 + i0}
~+~ \frac{1}{k^2 - \mu^{-1}k_0 + i0} \right)
 \frac{3 k_0^2 - 2 {\bf k}^2}{k^2 - 2 k_0},
\]

where the exchanged momentum $k$ is measured in units of the 
electron mass and the dimensionless parameter $\mu$ is given by the 
expression $\mu=m/(2M)$. Note that the expression in Eq.\ (\ref{polgen}) 
differs from the integral for the skeleton graph with two exchanged photons 
only by the substitution 

\begin{equation}
\frac{1}{k^2}\rightarrow\frac{\alpha}{\pi}
\int_0^1dv\frac{v^2(1-v^2/3)}{k^2(1-v^2)-4}.
\end{equation}

Let us now describe briefly the calculation of the contribution of relative 
order $\mu^2$ contained in Eq.\ (\ref{polgen}). First we combine the 
electron and the photon denominators, and obtain

\begin{equation} \label{derrep}
\delta E^{\text{pol}} ~=~ \alpha (Z\alpha) E_F~  \Bigl(-\frac{2}{\pi^2 \mu }
\Bigr) \int_0^1 \frac{dv}{1 - v^2}~v^2 \Bigl(1 - \frac{v^2}{3}\Bigr)
\int_0^1 {dz} 
\end{equation}
\[
(-\frac{\partial}{\partial a^2_p}) \int\frac{d^4 k}{i \pi^2}~  
\frac{1}{k^4 - \mu^{-2}  k_0^2 } 
\frac{3 k_0^2 - 2 {\bf k}^2}{(~-k^2 ~+~ 2b_p  k_0 ~+~ a^2_p ~)},
\]

where $a^2_p=4z/(1 - v^2)$, $b_p=1 - z$.

The general expression in Eq.\ (\ref{derrep}) contains nonrecoil radiative 
corrections of order $\alpha(Z\alpha)E_F$ \cite{kp,kks}, as well as 
radiative-recoil corrections of all orders in the electron-muon mass ratio 
generated by the polarization operator insertions and admits numerical 
calculation.  Due to the structure of the integrand explicit extraction of 
the corrections of definite order in the mass ratio is more involved. Direct 
application of the standard Feynman parameter methods leads to such 
integrals for the radiative-recoil corrections which do not admit expansion 
of the integrand over the small mass ratio before integration, making the 
analytic calculation virtually impossible. Instead of combining denominators 
in Eq.\ (\ref{derrep}) with the help of the Feynman parameters we use an 
approach developed earlier for calculation of the radiative-recoil 
corrections of order $\alpha(Z\alpha)(m/M)E_F$ (see, e.g., review in 
\cite{eksann1}). The idea is to perform integration over the exchanged 
momentum directly in spherical coordinates. Following this route we come 
to the expression

\begin{equation}     \label{polphi}
\delta E^{\text{pol}} ~=
\alpha (Z\alpha) E_F~  \Bigl(-\frac{4 \mu}{\pi^2 }\Bigr)
 \int_0^1 \frac{dv}{1 - v^2}~v^2 \Bigl(1 - \frac{v^2}{3}\Bigr)
 \int_0^1 {dz} 
\end{equation}
\[
 \frac{\partial}{\partial a^2_p} 
\Biggl\{\int_0^\infty {d k^2} ~
\frac{k^2 ( k^2 + a^2_p)}{( k^2 + a^2_p)^2 ~-~ 4 \mu^2 b^2_p k^4}
 \Bigl[~2\Phi_0 (k) ~+~ \Phi_1 (k) ~\Bigr]\Biggr\},
\]

which readily admits expansion in the small parameter $\mu$. We have 
introduced here standard auxiliary functions $\Phi_n(k)$, which emerge in 
all calculations in this paper. Definitions and properties of these 
functions are described in Appendix \ref{appa}.

The crucial property of the momentum integrand in Eq.\ (\ref{polphi}), which 
facilitates further  calculation, is that the denominator admits 
expansion in the small parameter $\mu$ prior to momentum integration. This 
is true due to the inequality  $4\mu^2b_p^2k^4/(k^2+a^2)^2\leq 4\mu^2$ which 
is obviously satisfied because $0\leq a_p< \infty$ and $0\leq b_p\leq 1$. In 
this way we may easily reproduce nonrecoil radiative corrections induced by 
the vacuum polarization insertions \cite{kp,kks}, and respective corrections 
of the first order in the mass ratio which were obtained in \cite{cp,ty}. We 
will omit here these calculations and describe only calculation of the 
corrections of order $(m/M)^2$. Expanding the denominator in Eq.\ 
(\ref{polphi}) in the small mass ratio we obtain

\begin{equation}      \label{polart}
\delta E^{\text{pol}} ~\approx
\alpha (Z\alpha) E_F~  \Bigl(-\frac{4 \mu}{\pi^2 }\Bigr)
 \int_0^1 \frac{dv}{1 - v^2}~v^2 \Bigl(1 - \frac{v^2}{3}\Bigr)
 \int_0^1 {dz} 
\end{equation}
\[
\frac{\partial}{\partial a^2_p} \Biggl\{\int_0^\infty {d k^2} ~k^2 
 \Biggl[~\frac{1}{ k^2 + a^2_p} ~+~ \frac{4 \mu^2 b^2_p k^4}{( k^2 + 
a^2_p)^3} \Biggr] ~ 
 \Bigl[~2\Phi_0^S 
(k) ~+~ 2\Phi_0^{\mu} (k) ~+~ \Phi_1^{\mu} (k) ~+~ 2\Phi_0^C (k) ~+~ 
\Phi_1^C (k) ~\Bigr]~\Biggr\}.  
\]

Functions $\Phi^C_i(k)$ do not depend on the small parameter $\mu$ (see 
Appendix \ref{appa}) and, hence, do not generate corrections of order 
$\mu^2$.  Calculation of the $\mu$-integrals (integrals with functions 
$\Phi^\mu$ in the integrands) is performed using an auxiliary parameter 
$\sigma$ ($1\ll\sigma\ll \mu^{-1}$) (see a more detailed discussion in the 
next Section \ref{electrsect}). Detailed analysis demonstrates that only 
the region of small momenta (in comparison with the inverse electron-muon 
mass ratio) generates corrections of the second order in $(m/M)^2$. 
All such corrections are connected with the product of the function 
$\Phi_0^\mu$ and of the term with factor $b_p$ in the numerator in Eq.\ 
(\ref{polart}) \footnote{All statements above are proved in Appendix 
\ref{appb}.} and we obtain

\begin{equation}          \label{totvac}
\delta E_{\text{vp}} ~=
\alpha (Z\alpha) E_F~  \frac{2}{\pi}
 \int_0^1 \frac{dv}{1 - v^2}~v^2 \Bigl(1 - \frac{v^2}{3}\Bigr)
 \int_0^1 \frac{dz}{a_p} 
15 \mu^2 b^2_p=\frac{3}{4}~
\alpha (Z\alpha)\biggl(\frac{m}{M}\biggr)^2 E_F.
\end{equation}

Besides the electron vacuum polarization one also has to consider the muon 
vacuum polarization insertion. The respective contribution differs from the 
one for the electron only by the substitution $k\rightarrow (M/m)k$ and has 
the form

\begin{equation}
\delta E_{\mu\text {vp}} ~= ~
\alpha (Z\alpha) E_F~  \frac{16 \mu}{\pi^3}
 \int_0^1 {dv}~v^2 \Bigl(1 - \frac{v^2}{3}\Bigr) 
\int_0^{\pi}{d \theta} \sin^2{\theta}~(2 + \cos^2{\theta}) ~
\end{equation}
\[
 \int_0^{\infty} {d k^2}~
\frac{1}{k^2 + 4 \cos^2 {\theta}} 
\frac{1}{4 + k^2(1 -v^2)} 
\frac{k^2}{k^2 + 16\mu^2 \cos^2 {\theta}}~.
\]

Representing the last factor in the integrand in the form

\begin{equation}  \label{identity}
\frac{k^2}{k^2 + 16\mu^2 \cos^2 {\theta}} ~\equiv ~
1 ~-~ \frac{16 \mu^2 \cos^2 {\theta}}{k^2 + 16\mu^2 \cos^2 {\theta}},
\end{equation}

we easily obtain

\begin{equation}
\delta E_{\mu \text {vp}} ~= ~\biggl(\frac{\pi^2}{3}-\frac{10}{9}\biggr)
\frac{\alpha (Z\alpha)}{\pi^2} \biggl(\frac{m}{M}\biggr)E_F~  
+O(\mu^3),  
\end{equation}

and, hence, the muon vacuum polarization does not generate contributions of 
order $(m/M)^2$. Hadron vacuum polarization insertions lead to the 
corrections of the same order as the muon vacuum polarization and will be 
ignored in our discussion.

\section{Electron-Line Contribution}                                               
\label{electrsect}

Radiative corrections to the hyperfine splitting 
induced by the radiative insertions in the electron line are given by the 
expression \cite{beks}

\begin{equation}      \label{electrgen} 
\delta E^{\text{e-line}} ~=~ \frac{Z\alpha}{\pi} E_F~ \Big(- \frac{3}{16\mu} 
\Big) ~ \int \frac{d^4 k}{i \pi^2}~ \frac{1}{(k^2 + i0)^2}~ \biggl( 
\frac{1}{k^2 + \mu^{-1}k_0 + i0} 
\end{equation} 
\[ ~+~ \frac{1}{k^2 - \mu^{-1}k_0 + 
i0}   \biggr) \langle \gamma^{\mu} \hat {k} \gamma^{\nu} \rangle _{(\mu)} ~ 
L_{\mu\nu}, 
\]

where the electron factor $L_{\mu\nu}$ is a 
sum of four terms $L_{\mu\nu}=\sum_1^4L^i_{\mu\nu}$ \cite{beks}, which after 
some transformations may be written in the form

\begin{equation}             \label{fermionfact}
L^{(1)}_{\mu \nu} = \frac{\alpha}{4\pi} 
\langle \gamma_{\mu} \hat {k} \gamma_{\nu}\rangle _{(e)} 
\int_0^1 dx  \int_0^x
\frac{dy}{(-k^2 + 2bk_0 + a^2)^3}~ 
\Big\{~c_{1}~  {\bf k}^2~+~c_{2}~  k^4 ~ \Big\},
\end{equation}
\[
L^{(2)}_{\mu \nu} = \frac{\alpha}{4\pi} 
\langle \gamma_{\mu} \hat {k} \gamma_{\nu}\rangle _{(e)} 
\int_0^1 dx  \int_0^x
\frac{dy}{(-k^2 + 2bk_0 + a^2)^2~}~ 
\Big\{~c_{3}~  k^2~+~2~c_{4}~  k_0~ \Big\},
\]
\[
L^{(3)}_{\mu \nu} = \frac{\alpha}{4\pi} 
\langle \gamma_{\mu} \gamma_{\nu}\rangle _{(e)} 
\int_0^1 dx  \int_0^x
\frac{dy}{(-k^2 + 2bk_0 + a^2)^2~}~ 
\Big\{~c_{5}~  k^2~+~2~c_{6}~  k^2~  k_0~ \Big\},
\]
\[
L^{(4)}_{\mu \nu} = \frac{\alpha}{4\pi} 
\langle \gamma_{\mu} \gamma_{\nu}\rangle _{(e)} 
\int_0^1 dx  \int_0^x
\frac{dy}{-k^2 + 2bk_0 + a^2}~ 
\Big\{~c_{7}~  k^2~ \Big\}.
\]

Auxiliary functions of the Feynman parameters $a(x,y)$ and $b(x,y)$ are 
defined by the relationships

\begin{equation}
a^2 ~=~ \frac{x^2}{y(1 - y)}, ~~~~~~ b ~=~ \frac{1 - x}{1 - y},
\end{equation}

and explicit expressions for the coefficient functions $c_i$ are collected 
in Table \ref{table1}.

Performing contractions of the products of the electron and muon spinor 
structures 

\begin{equation}
\langle \gamma_{\mu} \hat {k} \gamma_{\nu} \rangle _{(e)} 
\langle \gamma^{\mu} \hat {k} \gamma^{\nu} \rangle _{(\mu)}
~=~ \frac{8}{3}  (~-~3k_0^2 ~+~ 2{\bf k}^2~),
\end{equation}
\[
\langle \gamma_{\mu} \gamma_{\nu}\rangle _{(e)} 
\langle \gamma^{\mu} \hat {k} \gamma^{\nu}\rangle _{(\mu)}  ~=~ 8k_0,
\]

we arrive at the expression for the electron line contribution to 
the hyperfine splitting 

\begin{equation}      \label{electrcontr}
\delta E^{\text{e-line}} ~=~ \alpha (Z\alpha) E_F~  \frac{1}{8 \pi^2 \mu} ~
\int_0^1 {dx} \int_0^x {dy}
\int \frac{d^4 k}{i \pi^2}~ \frac{1}{(k^2 + i0)^2}~ 
\end{equation}
\[
\biggl(
\frac{1}{k^2 + \mu^{-1}k_0 + i0}
~+~ \frac{1}{k^2 - \mu^{-1}k_0 + i0}
\biggr) 
\]
\[
 \Bigl\{~ (~3k^2_0 ~-~ 2{\bf k}^2 ~) 
\Bigl[~\frac{c_{1}  {\bf k}^2 ~+~ c_{2}  (k^2)^2}
{(~-k^2 ~+~ 2bk_0 ~+~ a^2~)^3}
~+~\frac{ c_{3}  k^2 ~+~ c_{4}  2k_0}
{(~-k^2 ~+~ 2bk_0 ~+~ a^2~)^2} \Bigr]
\]
\[
~-~ 3k_0  \Bigl[
~\frac{c_{5}  k^2 ~+~ c_{6}  k^2  2k_0}
{(~-k^2 ~+~ 2bk_0 ~+~ a^2~)^2}
~+~ \frac{c_{7}  k^2}
{~-k^2 ~+~ 2bk_0 ~+~ a^2~} \Bigr] \Bigr\}\equiv\sum_1^7\delta E_i.
\]

Now we will discuss briefly calculation of the contributions to the 
hyperfine splitting of order $\alpha(Z\alpha)(m/M)^2E_F$ generated by the 
electron factor. Let us start our consideration with the simplest 
contribution corresponding to the coefficient function $c_7$ in Table 
\ref{table1}.  After Wick rotation and calculation of the angle integral we 
arrive at the expression

\begin{equation}
\delta E ~
=~ \alpha (Z\alpha) E_F~ 
\Bigl(-\frac{3 \mu}{ \pi^2}\Bigr)
 \int_0^1 {dx} \int_0^x {dy}~  b~  c_{7} ~
 \int_0^{\infty} {d k^2} ~
\frac{k^2}{~(k^2 + a^2)^2 ~-~ 4 \mu^2 b^2 k^4~} ~ 
\Phi_1 (k),
\end{equation}

where the auxiliary function $\Phi_1 (k)$ is defined in the Appendix 
\ref{appa}.

As in the case of the vacuum polarization discussed above one may expand the 
denominator of the integrand and obtain

\begin{equation}
\delta E ~\approx~ \alpha (Z\alpha) E_F~ 
\Bigl(-\frac{3 \mu}{ \pi^2}\Bigr)
 \int_0^1 {dx} \int_0^x {dy}~  b~  c_{7} ~
\end{equation}
\[
 \int_0^{\infty} {d k^2} ~
\frac{k^2}{(k^2 + a^2)^2}  
\biggl\{~1 ~+~ \frac{4~ \mu^2~ b^2 ~k^4}{(k^2 + a^2)^2}
\biggr\} 
\Bigl[\Phi_1^{\mu} (k) ~+~\Phi_1^C (k) ~ \Bigr],
\]

where the second term in the braces is at least of the second order in 
$\mu^2$ due to an obvious inequality  $4~ \mu^2~ b^2 ~k^4/(k^2 + a^2)^2 \leq 
4 \mu^2$. 

It is easy to see that the $c$-integral (integral which contains function 
$\Phi^C_1$ in the integrand) does not generate corrections of the 
second order in $\mu^2$ since the function  $\Phi^C_1$ is $\mu$-independent. 
Only the $\mu$-integral 

\begin{equation}              \label{muint7}
\delta E_7^\mu=\alpha (Z\alpha) E_F~ 
\Bigl(-\frac{3 \mu}{ \pi^2}\Bigr)
 \int_0^1 {dx} \int_0^x {dy}~  b~  c_{7}
\int_0^{\infty} {d k^2} ~
\frac{k^2}{(k^2 + a^2)^2} \Phi_1^{\mu} (k),
\end{equation}
       
generates corrections of the second order in the mass ratio.

As we already mentioned in Section \ref{polarsect} the integrals of this 
kind are usually calculated with the help of an auxiliary parameter 
$\sigma$, which satisfies the inequality $1\ll \sigma\ll \mu^{-1}$. 
The parameter $\sigma$ is used to separate the momentum integration into two 
regions, a region of small momenta $0\leq k\leq\sigma$, and a region of 
large momenta $\sigma\leq k<\infty$. In the region of small momenta one uses 
for simplification of the integrand the condition $\mu k\ll 1$, and in the 
region of large momenta the condition $k\gg 1$. Note that for $k\simeq 
\sigma$ both conditions on the integration momenta are valid simultaneously, 
so in the sum of the low-momenta and high-momenta integrals all 
$\sigma$-dependent terms cancel and one obtains a $\sigma$-independent 
result for the total momentum integral (see more detailed discussion of this 
method, e.g., in \cite{eksann1}).  Sometimes one needs to use more stringent 
restrictions on the auxiliary parameter $\sigma$ in order to obtain a result 
with higher accuracy in $\mu$.  This case, as well as the sample 
calculation of the $\mu$-integral for the $\delta E_7$ contribution are 
discussed in more detail in Appendix \ref{appb}. The analysis performed in 
this Appendix leads to the conclusion that all corrections of order 
$(m/M)^2$ are generated by the low-momentum part of the $\mu$-integral 
in Eq.\ (\ref{muint7}), and in order to obtain these corrections one simply 
has to extract that contribution of order $\mu^2$ to this integral which 
remains finite when the auxiliary parameter $\sigma$ goes to infinity.  Then 
it is easy to obtain the respective contribution 

\begin{equation}  \label{e7}
\delta E_{7} = \alpha (Z\alpha) E_F~ 
\Bigl(~-~\frac{9 \mu^2}{2 \pi}~\Bigr) 
\int_0^1 {dx} \int_0^x {dy}~  b~  c_{7}~  a ~
=-\frac{9}{32}\alpha (Z\alpha)\biggl(\frac{m}{M}\biggr)^2 E_F.
\end{equation}

Calculation of the contribution to the hyperfine splitting induced by the 
term with the coefficient function $c_5$ in Eq.\ (\ref{electrcontr}) 
proceeds essentially along the same lines as in the case of the $\delta E_7$ 
contribution. Again the $c$-integral does not generate corrections of order 
$\mu^2$ to the hyperfine splitting. An integral representation for the 
 $\mu$-integral contribution may  be obtained from the representation in 
Eq.\ (\ref{muint7}) by the substitution $c_7\rightarrow 
-c_5(\partial/(\partial a^2))$

\begin{equation}  \label{e5muint}
\delta E_5^\mu 
~=~ \alpha (Z\alpha) E_F~ 
\Bigl(-\frac{6 \mu}{ \pi^2}\Bigr)
 \int_0^1 {dx} \int_0^x {dy}~  b~  c_{5} ~
~ \int_0^{\infty} {d k^2} ~
\frac{k^2}{(k^2 + a^2)^3}   ~\Phi_1^{\mu} (k).
\end{equation}

Correction of order $\mu^2$ to the hyperfine splitting is generated by the 
term linear in $\mu$ in the low momentum expansion of the function 
$\Phi_1^{\mu} (k)$. Due to a higher power of the denominator in the 
integrand in Eq.\ (\ref{e5muint}) than in Eq.\ (\ref{muint7}) the respective 
momentum integral is ultraviolet convergent and we immediately obtain

\begin{equation}         \label{e5}
\delta E_{5} 
=~ \alpha (Z\alpha) E_F~  \frac{9 \mu^2}{4 \pi} 
\int_0^1 {dx} \int_0^x {dy}~  \frac{b~  c_{5}}{a} ~
=-~\frac{3}{32}\alpha (Z\alpha) \biggl(\frac{m}{M}\biggr)^2E_F.
\end{equation}

In the case of the weight function $c_6$ in Eq.\ (\ref{electrcontr}) the 
$\mu$-integral has the form

\begin{equation}
\delta E_{6}^\mu = \alpha (Z\alpha) E_F~ 
\frac{3 \mu}{\pi^2}
 \int_0^1 {dx} \int_0^x {dy}~  c_{6} ~
\int_0^{\infty} {d k^2} ~
\frac{k^2}{(k^2 + a^2)^2}   ~\Phi_1^{\mu} (k) .
\end{equation}

Again extracting the term linear in $\mu$ from the integrand we obtain

\begin{equation}                  \label{e6}
\delta E_{6} = \alpha (Z\alpha) E_F~ 
\frac{9 \mu^2}{2 \pi} 
\int_0^1 {dx} \int_0^x {dy}~  c_{6}~  a ~
=-\frac{3}{32}\alpha (Z\alpha)\biggl(\frac{m}{M}\biggr)^2 E_F.
\end{equation}

Unlike all cases discussed above the $\mu$-integral for the weight function 
$c_4$ depends not only on the function $\Phi^\mu_{1}(k)$ but also on the 
function $\Phi^\mu_{2}(k)$ 

\begin{equation}                \label{mu4phi}
\delta E_{4}^\mu = \alpha (Z\alpha) E_F~ 
\frac{4 \mu}{\pi^2}
 \int_0^1 {dx} \int_0^x {dy}~  b~  c_{4} ~
\int_0^{\infty} {d k^2} ~
\frac{k^2}{(k^2 + a^2)^3}   \Bigl[~2 \Phi_1^{\mu} (k) ~
~+~ \Phi_2^{\mu} (k)~\Bigr] .
\end{equation}

However, the conclusion of Appendix \ref{appb} that only the low-momentum 
part of the $\mu$-integrals leads to corrections of order $\mu^2$ to 
hyperfine splitting remains valid, even in this case. We may repeat the 
considerations of Appendix \ref{appb} for this case, and check that the 
additional high-momentum contributions to the integral generated by the 
auxiliary high momentum integrals $V_{202}$ and $V_{212}$ (see definitions 
of these integrals and discussion of their properties in Appendix 
\ref{appc}) differ from the respective contributions $V_{201}$ and $V_{211}$ 
connected with the function $\Phi^\mu_{1}(k)$ only by the factor $1/4$. 
Hence they do not lead to corrections of order $\mu^2$ to the hyperfine 
splitting. Moreover, terms linear in $\mu$ are missing in the small 
momentum expansion of the function $\Phi^\mu_{2}(k)$. Hence, one may 
completely ignore presence of the function $\Phi^\mu_{2}(k)$ in the 
integrand in Eq.\ (\ref{mu4phi}) during calculation of the contribution to 
the hyperfine splitting of relative order $\mu^2$.  Then calculation of the 
contribution $\delta E_4$ may be done exactly in the same way as calculation 
of the contribution $\delta E_5$ above and one obtains

\begin{equation}                           \label{e4}
\delta E_{4} 
=~ \alpha (Z\alpha) E_F~  \Bigl(-\frac{3 \mu^2}{\pi}\Bigr) 
\int_0^1 {dx} \int_0^x {dy}  \frac{b~  c_{4}}{a} 
=\frac{3}{16}\alpha (Z\alpha)\biggl(\frac{m}{M}\biggr)^2 E_F.
\end{equation}

We will combine consideration of the contributions to hyperfine 
corresponding to the weight factors $c_1$, $c_2$, and $c_3$.
In this case, after the angular integration, we arrive at the expression

\begin{equation}
\delta E 
=\alpha (Z\alpha) E_F~  \frac{ \mu}{4\pi^2}
\int_0^1 {dx} \int_0^x {dy}
\int_0^{\infty} {d k^2} k^2 
\Biggl\{ \Bigl[~(c_1 + c_2  k^2) 
\Bigl(\frac{\partial}{\partial a^2}\Bigr)^2 +
2c_3  \frac{\partial}{\partial a^2} \Bigr] 
\end{equation}
\[
\frac{ k^2 + a^2}{~(k^2 + a^2)^2 ~-~ 4 \mu^2 b^2 k^4 ~}
\Bigl[2 \Phi_0 (k) + \Phi_1 (k) \Bigr]
- c_1 \Bigl(\frac{\partial}{\partial a^2}\Bigr)^2 
\frac{ k^2 + a^2}{(k^2 + a^2)^2 - 4 \mu^2 b^2 k^4 }
\Bigl[2 \Phi_1 (k) + \Phi_2 (k)  \Bigr] \Biggr\}.
\]

Differentiation and expansion in a series in $\mu^2$ leads to the 
integral

\begin{equation}
\delta E \approx
\alpha (Z\alpha) E_F  \frac{ \mu}{2\pi^2}
\int_0^1 {dx} \int_0^x {dy} 
\int_0^{\infty} {d k^2}  k^2   
\Biggl\{\Bigl[\frac{c_1 + c_2  k^2}{(k^2 + a^2)^3} -
\frac{ c_3}{(k^2 + a^2)^2} \Bigr]
\end{equation}
\[
\Bigl[2 \Phi_0 (k) + \Phi_1 (k)  \Bigr]+12 \mu^2 b^2 k^4 
\Bigl[\frac{2 c_1 + 2 c_2  k^2}{(k^2 + a^2)^5} -
\frac{ c_3}{(k^2 + a^2)^4} \Bigr] 
\Bigl[2 \Phi_0 (k) + \Phi_1 (k)  \Bigr]
\]
\[
- c_1 
\Bigl[\frac{1}{(k^2 + a^2)^3} +
\frac{24 \mu^2 b^2 k^4}{(k^2 + a^2)^5} \Bigr] 
\Bigl[2 \Phi_1 (k) + \Phi_2 (k)  \Bigr] \Biggr\}.
\]

As usual we write the functions $\Phi_i(k)$ as sums of the functions 
$\Phi^\mu_i(k)$ and $\Phi^C_i(k)$ (see Appendix \ref{appb}). Let us
mention here that the classical nonrecoil electron-factor 
contribution to the hyperfine splitting \cite{kp,kks} is produced by the 
term with $\Phi_0^S=2/(\mu k)$ in the integral above

\begin{equation}
\delta E_{KP} ~=~
\alpha (Z\alpha) E_F~  \frac{2}{\pi^2}
 \int_0^1 {dx} \int_0^x {dy} ~
\int_0^{\infty} {d k} ~ k^2 ~
\Bigl[~\frac{c_1 + c_2  k^2}{(k^2 + a^2)^3} ~-~
\frac{ c_3}{(k^2 + a^2)^2} \Bigr]
\end{equation}
\[
=~ \alpha (Z\alpha) E_F~  \Bigl(~\ln2 ~-~ \frac{13}{4}~\Bigr).
\]

Contributions of relative order $\mu^2$ are connected only with 
the functions $\Phi_i^\mu$, and are hidden in the integral

\begin{equation}            \label{e123gen}
\delta E_{123}^\mu \approx
\alpha (Z\alpha) E_F  \frac{\mu}{2\pi^2}
 \int_0^1 {dx} \int_0^x {dy} 
\int_0^{\infty} {d k^2}  k^2 
\Biggl\{
 \Bigl[2\Phi_0^{\mu} (k) + \Phi_1^{\mu} (k)\Bigr] 
\Bigl[\frac{c_1 + c_2  k^2}{(k^2 + a^2)^3} -
\end{equation}
\[
\frac{ c_3}{(k^2 + a^2)^2} \Bigr]
+24 \mu^2 b^2 k^4 
\Phi_0^S (k) 
\Bigl[\frac{2 c_1 + 2 c_2  k^2}{(k^2 + a^2)^5} -
\frac{ c_3}{(k^2 + a^2)^4} \Bigr] 
- \Bigl[2 \Phi_1^{\mu} (k) + \Phi_2^{\mu} (k)  \Bigr]
 \frac{c_1}{(k^2 + a^2)^3}   \Biggr\}.
\]

As usual, all contributions of relative order $\mu^2$ are generated by the 
low-momentum integration region $0\leq k\leq\sigma$, where one can use 
the small momentum approximation for the functions $\Phi^\mu_i$

\begin{equation}
2\Phi_0^{\mu} (k) ~+~ \Phi_1^{\mu} (k) ~\approx ~
- \frac{3}{2} ~+~ (\mu k )^2 , 
\end{equation}
\[
2\Phi_1^{\mu} (k) ~+~ \Phi_2^{\mu} (k) ~\approx ~
\frac{9}{8} ~-~ 2 \mu k ~+~ \frac{3}{2} (\mu k )^2 .
\]

Due to absence of terms linear in $\mu$  in the low-momentum 
expansion of the function $2\Phi_0^{\mu} (k)+\Phi_1^{\mu} (k)$, 
the respective term  in the integrand in Eq.\ (\ref{e123gen}) does not 
generate corrections of relative order $\mu^2$. Hence all such corrections 
in Eq.\ (\ref{e123gen}) are connected with the functions $\Phi_0^S (k)= 
1/\mu k$ and $2\Phi_1^{\mu} (k)+ \Phi_2^{\mu} (k)\approx- 2 \mu k $, and 
are given by the expression

\begin{equation}                                    \label{e123}
\delta E_{123} =
\alpha (Z\alpha) E_F~  \frac{2\mu^2}{\pi^2}
 \int_0^1 {dx} \int_0^x {dy} ~
\int_0^{\infty} {d k} ~  \Biggl\{ ~
\frac{c_1  k^4}{(k^2 + a^2)^3} ~+~
\end{equation}
\[
12 b^2 k^6 ~
\Bigl[~\frac{2 c_1 + 2 c_2  k^2}{(k^2 + a^2)^5} ~-~
\frac{ c_3}{(k^2 + a^2)^4} ~\Bigr]~\Biggr\} 
=
(-6 \ln2 - \frac{39}{32})\alpha (Z\alpha)\biggl(\frac{m}{M}\biggr)^2 E_F.
\]

Now we are in a position to write down the total contribution to the 
hyperfine splitting of relative order $\mu^2$ generated by the electron 
factor, which is given by the sum of the contributions in 
Eq.\ (\ref{e7}), Eq.\ (\ref{e5}), Eq.\ (\ref{e6}), Eq.\ (\ref{e4}), and Eq.\ 
(\ref{e123}) (see also Table \ref{table2})

\begin{equation}   \label{electr}
\delta E_{\text{ el}}=(-6  \ln2 ~-~ \frac{3}{2}~)
\alpha (Z\alpha) \biggl(\frac{m}{M}\biggr)^2E_F.
\end{equation}

\section{Muon-Line Contribution}                                               

After an obvious substitution $\alpha\rightarrow Z^2\alpha$
the electron-line factor in Eq.\ (\ref{fermionfact})  properly  describes 
radiative corrections to the muon line as well.  Then the total expression 
for the muon-line contribution to hyperfine splitting of order 
$(Z^2\alpha)(Z\alpha)E_F$ may be easily obtained from Eq.\ (\ref{electrgen}) 
and has the form

\begin{equation}                \label{muongen}
\delta E^{\mu-\text{line}} ~=~ (Z^2 \alpha)(Z \alpha) E_F~ 
\frac{\mu}{2 \pi^2} ~ \int_0^1 {dx} \int_0^x {dy}
\int \frac{d^4 k}{i \pi^2}~\frac{1}{(k^2 + i0)^2}~ 
\biggl( \frac{1}{k^2 + 4\mu k_0 + i0}
\end{equation}
\[
~+~ \frac{1}{k^2 - 4\mu k_0 + i0} \biggr) 
 \Bigl\{~(~3k^2_0 ~-~ 2{\bf k}^2 ~)  
\Bigl[~\frac{c_{1}  {\bf k}^2
~+~ c_{2}  (k^2)^2} {(~-k^2 ~+~ 2bk_0 ~+~ a^2~)^3} ~+~\frac{ c_{3}
 k^2 ~+~ c_{4}  2 k_0} {(~-k^2 ~+~ 2bk_0 ~+~ a^2~)^2} \Bigr]
\]
\[
~-~ 3 k_0  \Bigl[ ~\frac{c_{5}  k^2 ~+~
c_{6}  k^2  2 k_0} {(~-k^2 ~+~ 2bk_0 ~+~ a^2~)^2} ~+~ \frac{c_{7}
 k^2} {~-k^2 ~+~ 2bk_0 ~+~ a^2~} \Bigr] \Bigr\}.
\]

Apparent distinctions between Eq.\ (\ref{electrgen}) and Eq.\ 
(\ref{muongen}) remind us that while in Eq.\ (\ref{electrgen}) the 
dimensionless integration momentum is measured in the units of electron 
mass, in Eq.\ (\ref{muongen}) it is measured in the muon mass units.

After standard transformations the integral in Eq.\ (\ref{muongen}) acquires 
the form

\begin{equation}     \label{muongen1}        
\delta E^{\mu-\text{line}} ~=~  (Z^2 \alpha)(Z \alpha) E_F~ 
\frac{\mu}{\pi^3} ~ 
\int_0^{\pi} {d \theta}~ \sin^2 {\theta} \int_0^{\infty} {d k^2} 
 \Biggl(~1 ~-~ \frac{16\mu^2 ~\cos^2 {\theta}}{k^2 
+ 16\mu^2 ~\cos^2 {\theta}} ~ \Biggr)  I(k),
\end{equation}

where the first factor in the integrand is simplified with the help of 
the identity in Eq.\ (\ref{identity}), and the second factor is given by the 
integral

\begin{equation}        \label{iint}
I(k) ~=~ \int_0^1 {dx} \int_0^x {dy} 
\Bigl\{~(2 + \cos^2{\theta})  \Bigl[~(c_1 \sin^2{\theta} ~+~ c_2 k^2)
 \Bigl(\frac{\partial}{\partial a^2}\Bigr)^2 ~+~
2 c_3  \frac{\partial}{\partial a^2}~ \Bigr] (k^2 + a^2)
\end{equation}
\[
-~ 8 b c_4 (2 + \cos^2{\theta}) \cos^2{\theta} 
 \frac{\partial}{\partial a^2} ~+~
12 b \cos^2{\theta} \Bigl(~c_5  \frac{\partial}{\partial a^2} 
~-~c_7 \Bigr)
\]
\[
-~ 12 c_6  \cos^2{\theta}  \frac{\partial}{\partial a^2}
(k^2 + a^2)~ \Bigr\}~
\frac{1}{~(k^2 + a^2)^2 ~+~ 4 b^2 k^2 \cos^2{\theta}~}.
\]

The first factor in the parenthesis in the integrand in Eq.\ 
(\ref{muongen1}) generates only corrections of relative order $\mu$, since 
the integral $I(k)$ is $\mu$-independent. The respective correction to 
hyperfine splitting was calculated long time ago \cite{sty,eks6,eks7}, and 
we will not consider it here. Correction of order $\mu^2$ in Eq.\ 
(\ref{muongen1}) could be connected only with second term in the parenthesis 
which apparently generates a contribution of order $\mu^3$. However, the 
integral $I(k)$ behaves like $1/k$ at small momenta, and this singular 
behavior reduces the apparent power of $\mu$, generating a correction of 
relative order $\mu^2$.  In order to obtain this correction we have to 
calculate the leading term in the low-momentum expansion of the integral 
$I(k)$.  

Let us outline briefly the respective calculation. First, we note that the 
most singular $k$ behavior of the integral in Eq.\ (\ref{iint}) is 
connected with the region of small values of the Feynman parameter $x$. 
Hence, in the limit of small $k$ and $x$ we preserve in the integrand only 
nonvanishing terms. The calculation is facilitated by the change of variable 
$y=xz$.  Then use the following approximations

\begin{equation}
a^2\approx\frac{x}{z},\qquad b\approx 1,\qquad k^2+a^2\approx \frac{x}{z},
\end{equation}
\[
c_1 ~\approx ~ 16~  (~\frac{1}{z} - 3 - 2\ln{x}) , ~~~~~~~~
c_3 ~\approx ~ \frac{1}{x}~  (~\frac{1}{z} - 26 + 6z - 16\ln{x}) , 
\]
\[            
c_4 ~\approx ~ \frac{2}{z} - 5,~~~~
c_5 ~\approx ~ \frac{6}{z} - 8,~~~~
c_6 ~\approx ~ - \frac{1-z}{x},~~~~
c_7 ~\approx ~  \frac{2}{x}.
\]

The term with the coefficient function $c_2$ in Eq.\ (\ref{iint}) may be 
omitted since it is explicitly multiplied by the small factor $k^2$. Then 
the leading infrared contribution to the function $I(k)$ may be written as

\begin{equation}
I(k) ~\approx ~ 
\int_0^1 {dz}  z^2 \int_0^1 {dx} 
\Bigl\{~(2 + \cos^2{\theta})  
\Bigl[~16 \sin^2{\theta}  z 
(~\frac1z - 3 - 2\ln{x}) 
x \Bigl(\frac{\partial}{\partial x}\Bigr)^2 x 
\end{equation}
\[
+~ 2 (\frac1z - 26 + 6z - 16\ln{x})  
\frac{\partial}{\partial x} x~ \Bigr] ~+~
8 (2 + \cos^2{\theta})  \cos^2{\theta}  (- 2 + 5z)
 x \frac{\partial}{\partial x}  
\]
\[
+~ 12 \cos^2{\theta}  \Bigl[~(6 - 8z)
 x \frac{\partial}{\partial x} ~-~ 2 ~\Bigr] ~+~ 
12 \cos^2{\theta}  (1 - z)
 \frac{\partial}{\partial x} x ~ \Bigr\}                  
\frac{1}{x^2 ~+~ 4 k^2 z^2 \cos^2{\theta}}.
\]

Integrating by parts we reduce all infrared-hard terms to the standard 
integral

\begin{equation}
\int_0^1 \frac{dx}{x^2 ~+~ 4 k^2 z^2 \cos^2{\theta}} ~\approx ~
\frac{\pi}{4 k z \cos{\theta}} ,
\end{equation}

and arrive at the relationship

\begin{equation}
I(k) ~\approx ~ 
\int_0^1 {dz}~  z^2 
\Bigl[~32  (2 + \cos^2{\theta})  (1 - z \sin^2{\theta}) 
~+~ 8 (2 + \cos^2{\theta})  \cos^2{\theta}  (2 - 5z)
\end{equation}
\[
-~ 12 \cos^2{\theta}  (6 - 8z + 2) ~ \Bigr] 
\frac{\pi}{4 k z \cos{\theta}}~= ~
\frac{4 \pi}{3}  \frac{2 + \cos^4{\theta}}{k \cos{\theta}}.
\]

Now we easily obtain radiative-recoil correction of the relative order 
$\mu^2$ from the second term in the integrand in Eq.\ (\ref{muongen1})

\begin{equation}       \label{totmuon}
\delta E_{\text{ml}} ~=~ - \frac{17}{12}~ 
(Z^2 \alpha) (Z\alpha) \biggl(\frac{m}{M} \biggr)^2  E_F.
\end{equation}

\section{Discussion of Results}

Let us consider first the electron-line contribution in Eq.\ (\ref{electr}). 
This result demonstrates that the discrepancy between the results in 
\cite{sty} and in \cite{eks5,beks} for the radiative-recoil 
corrections induced by the electron line cannot be explained by the 
contributions of the terms of higher  order in the mass ratio to the result 
in \cite{sty}. Numerically we obtain from Eq.\ (\ref{electr})

\begin{equation}
\delta E_{\text{el}}\approx-0.03 \mbox{kHz},
\end{equation}

which is too small to explain the difference $0.22$ kHz between the results 
in \cite{sty} and \cite{eks5,beks}. We would like to remind the 
reader that the coinciding results in \cite{eks5} and 
\cite{beks} were obtained completely independently in different gauges, so 
there is little, if any, doubt that this result is correct.

The total one-loop radiative-recoil correction of relative order $(m/M)^2$ 
is given by the sum of the results in Eq.\ (\ref{totvac}), Eq.\ 
(\ref{electr}), and Eq.\ (\ref{totmuon})

\begin{equation}          \label{total}
\delta E_{\text{rad-rec}}=
\left[(-6  \ln2 ~-~ \frac{3}{4}~)~
\alpha (Z\alpha) 
- \frac{17}{12}~ 
(Z^2 \alpha) (Z\alpha)\right] \biggl(\frac{m}{M} \biggr)^2  E_F,
\end{equation}

or, recalling that $Z=1$ for muonium

\begin{equation}
\delta E_{\text{rad-rec}}=
(-6  \ln2 ~-~ \frac{13}{6}~)~
\alpha^2 \biggl(\frac{m}{M}\biggr)^2E_F
\approx-6.32555\ldots\alpha^2 \biggl(\frac{m}{M}\biggr)^2E_F.
\end{equation}

Let us mention a minor subtlety connected with the definition of the Fermi 
energy in Eq.\ (\ref{total}).  As is well known, there are two common 
definitions for the Fermi energy: one includes the muon anomalous magnetic 
moment, and the other does not. Both definitions are useful in presenting 
results for the high order corrections to the hyperfine splitting. 
Corrections, generated at the scale of order $1/m$ or larger (small 
intermediate momenta) are multiplied by the Fermi energy which includes the 
muon anomalous momentum.  But for the corrections generated at the scale 
comparable with the muon Compton length $1/M$ the muon anomalous momentum 
cannot be factorized, and such corrections are multiplied by the Fermi 
energy without the muon anomalous moment. As we have seen above, 
radiative-recoil corrections of the second order in mass ratio originate 
from the distances larger than the electron Compton length, and, hence, the 
Fermi energy in Eq.\ (\ref{total}) should include the muon anomalous 
moment. In this way we properly take into account some corrections of 
higher order than $\alpha(Z\alpha)(m/M)E_F$ but due to the smallness of the 
correction under consideration this is irrelevant from the purely 
phenomenological point of view.

Numerically the contribution to hyperfine splitting obtained in this paper 
is equal to

\begin{equation}
\delta E_{\text{rad-rec}}=-0.0351\ldots \mbox{kHz},
\end{equation}

and should be taken into account, along with some other small corrections to 
hyperfine splitting calculated recently, when comparing  the pending 
experimental results with the theory.

\acknowledgements
M. E. and V.S. are deeply grateful for the kind hospitality of the Physics 
Department at Penn State University, where this work was performed. The 
authors appreciate the support of this work by the National Science 
Foundation under grant number PHY-9421408.

\appendix

\section{Standard auxiliary functions}
\label{appa}

All contributions to hyperfine splitting in the main body of this paper 
are written in terms of the auxiliary functions $~\Phi_n (k) ~$ 
($n=0,1,2,3$) defined by the relationship

\begin{equation}
\Phi_n (k) ~ \equiv ~  \frac{1}{\pi \mu^2}
\int_0^{\pi} {d \theta}~ \sin^2 {\theta}   \cos^{2n} {\theta}
~
\frac{~(k^2 + a^2)^2 ~-~ 4~ \mu^2~ b^2~ k^4~}
{\Bigr(~k^2 + \mu^{-2} ~\cos^2 {\theta} ~ \Bigl) 
\Bigl[~(k^2 + a^2)^2 ~+~ 4~ b^2~ k^2 ~\cos^2{\theta}~\Bigr]}.
\end{equation}

The integral over angles may be explicitly calculated, and the result of the 
integration is conveniently written as a sum 

\begin{equation}
\Phi_n (k) ~ \equiv ~ \Phi_n^S (k) ~+~\Phi_n^{\mu} (k) ~+~\Phi_n^C (k),
\end{equation}

where

\begin{equation}
\Phi_n^S (k)  ~=~ \frac{\delta_{n0}}{\mu k} ,
\end{equation}
\begin{equation}
\Phi_0^{\mu} (k)  ~=~ W(\xi_{\mu}) ~-~ \frac{1}{\sqrt{\xi_{\mu}}},
\end{equation}
\begin{equation}
\Phi_1^{\mu} (k)  ~=~ -~\xi_{\mu}  W(\xi_{\mu}) ~+~ \frac{1}{2},
\end{equation}
\begin{equation}
\Phi_2^{\mu} (k)  ~=~ \xi_{\mu}  \Bigl(~\xi_{\mu}  W(\xi_{\mu})
~-~ \frac{1}{2} \Bigr) ~+~ \frac{1}{8} ,
\end{equation}
\begin{equation}
\Phi_3^{\mu} (k)  ~=~ -~ \xi_{\mu}  \Bigl[~\xi_{\mu} 
\Bigl(~\xi_{\mu}  W(\xi_{\mu}) ~-~ \frac{1}{2} \Bigr) ~+~
\frac{1}{8}~\Bigr] ~+~ \frac {1}{16} ,
\end{equation}
\begin{equation}
\Phi_0^C (k)  ~=~-~ W(\xi_C) ,
\end{equation}
\begin{equation}
\Phi_1^C (k)  ~=~ \xi_C  W(\xi_C) ~-~ \frac{1}{2},
\end{equation}
\begin{equation}
\Phi_2^C (k)  ~=~-~\xi_C  \Bigl(~\xi_C  W(\xi_C)
~-~ \frac{1}{2} \Bigr) ~-~ \frac{1}{8} ,
\end{equation}
\begin{equation}
\Phi_3^C (k)  ~=~ \xi_C  \Bigl[~\xi_C 
\Bigl(~\xi_C  W(\xi_C) ~-~ \frac{1}{2} \Bigr) ~+~
\frac{1}{8}~\Bigr] ~-~ \frac{1}{16} .
\end{equation}

The standard function $W(\xi)$ has the form
 
\begin{equation}
W(\xi) ~=~ \sqrt {1 ~+~ \frac{1}{\xi}} ~-~ 1 .
\end{equation}

and 

\begin{equation}
\xi_{\mu} ~=~ \mu^2 ~ k^2 ,~~~~~
\xi_C ~=~ \frac{(k^2 + a^2)^2}{4 b^2 k^2}.
\end{equation}

One may easily obtain asymptotic expressions for the function $W(\xi)$

\begin{equation}
\lim_{\xi\rightarrow 0}W(\xi)\rightarrow\frac{1}{\sqrt{\xi}},
\end{equation}
\[
\lim_{ \xi\rightarrow \infty}W(\xi)\rightarrow\frac{1}{2\xi}.
\]

High- and low-momentum asymptotic 
expressions for the functions $\Phi_i(k)$ also may be easily calculated. 
Let us cite low-momentum expansions, which where used in the main text for 
calculation of the contributions to the hyperfine splitting of relative 
order $\mu^2$

\begin{equation}
\Phi_0^{\mu} (k) ~\approx ~  -~1 ~+~ \frac{\mu k}{2} ,
\end{equation}
\begin{equation}
\Phi_1^{\mu} (k) ~\approx ~  \frac{1}{2} ~ - ~  \mu k  ~ + ~
(\mu k )^2,
\end{equation}
\begin{equation}
\Phi_2^{\mu} (k) ~\approx ~  \frac{1}{8} ~ - ~ \frac{(\mu k)^2}{2} .
\end{equation}

\section{Calculation of $\mu$-integrals}
\label{appb}

In this Appendix we will consider the main tricks used in calculation of the 
$\mu$-integrals. Consider as an example calculation of the integral relevant 
for the $\delta E_7$ correction

\begin{equation}  \label{mu7}
F(\mu) 
=\int_0^1 {dx} \int_0^x {dy}~  b~  c_{7}
\int_0^{\infty} {d k^2} ~
\frac{k^2}{(k^2 + a^2)^2} \Phi_1^{\mu} (k).
\end{equation}

We need to calculate contributions order $\mu$ to this integral.  As was 
discussed in the main text we introduce an auxiliary parameter $\sigma$ 
($1\ll\sigma\ll\mu^{-1}$) and consider separately contributions of the large 
and small integration momenta

\begin{equation}
F(\mu)=F^<(\mu)+F^>(\mu).
\end{equation}

In the large momenta region we would like to use the relationship 
$k\geq\sigma\gg1$ in order to simplify the integrand before integration over 
momenta. The simplest approach would be to expand the denominator in a power 
series over $a^2/k^2$ prior to integration over the Feynman parameters $x$ 
and $y$. However, it is easy to see that the second term of this naive 
expansion would lead to a divergence in the integration over $y$ when $y$ 
goes to zero. Of course, the initial integral is convergent, and this 
divergence is a  result of an improper expansion. A safe approach would be 
to calculate first the integrals over the Feynman parameters, and only then 
to expand in large $k^2$. Emergence of the logarithm of $k^2$ in this 
expansion corresponds to the divergence which was encountered in attempting 
to make the large $k^2$ expansion prior to the Feynman parameter 
integration.

However, we would like to avoid tedious calculation of  all  integrals 
over $x$ and $y$ prior to the large momentum expansion. As we will 
demonstrate now this may be achieved with the help of a trick, 
which will significantly simplify further calculations. The idea is to use  
the following representation for the coefficient function in Eq.\ 
(\ref{mu7})

\begin{equation}  \label{tilderepr}
bc_7=\widetilde{bc_7}+(bc_7-\widetilde{bc_7}),
\end{equation}

where 

\begin{equation}
\widetilde{bc_7}=(1-2y)(b(x,y)c_7(x,y))|_{y=0}.
\end{equation}

The weight function $(bc_7-\widetilde{bc_7})$ is obviously proportional to 
$y$, and, hence, the  high-momenta integrals with such weight function admit 
naive expansion over $1/k^2$ prior to integration over the Feynman 
parameters. Integration over $y$ in the integral with the weight function 
$\widetilde{bc_7}$ is simplified due to the presence of the factor 
$1-2y=[y(1-2y)]'$.

Let us consider first the integral with the weight function 
$\widetilde{bc_7}$

\begin{equation}   \label{fmuint}
\widetilde{F}(\mu) 
=2\int_0^1 {dx} \int_0^x {dy} \frac{(1-x)^2}{x}(1-2y)
\int_0^{\infty} {d k^2} ~
\frac{k^2}{(k^2 + a^2)^2} \Phi_1^{\mu} (k).
\end{equation}

In order to calculate the high-momenta contribution we first integrate over 
variable $y$, expand the result in the series over $1/k^2$, and then 
integrate over $x$

\begin{equation}  
\widetilde{F}^>(\mu) 
=
2\int_{\sigma}^{\infty} {d k^2} 
\biggl\{~\frac{1}{4 k^2} ~-~ \frac{1}{3 k^4}
 \ln{k} \biggr\}  ~\Phi_1^{\mu} (k).
\end{equation}

Note that we took into account only the first two terms in the expansion 
over $1/k^2$. As we will see below, due to the special choice of the 
auxiliary parameter $\sigma$, these terms are sufficient for calculation of 
the integral above with accuracy linear in $\mu$. The remaining 
momentum integration is easily performed with the help of the standard 
integrals $V_{kmn}$ which are calculated in Appendix \ref{appc}, and we 
obtain

\begin{equation}  
\widetilde{F}^>(\mu) 
=2
\Bigl(~\frac{1}{4}  V_{101}
-\frac{1}{3}  V_{211} ~ \Bigr) ~\approx
2  \Bigl[~-~\frac{1}{4}  \ln{(2 \mu)}
~-~ \frac{1}{8} ~-~ \frac{1}{4}  \ln{\sigma} 
\end{equation}
\[
+~ \frac{\mu \sigma}{2} ~-~ \frac{(\mu \sigma )^2}{4}
~-~ \frac{1}{\sigma^2}  \Bigl(~ \frac{1}{6}  \ln{\sigma}
~+~ \frac{1}{12}~ \Bigr)~\Bigr].
\]

The result above is a power series in the small parameters $(\mu\sigma)$ and 
$1/\sigma^2$. In order to get rid of higher terms in the expansion we 
further specify  the magnitude of the auxiliary parameter $\sigma\simeq 
1/\sqrt\mu$. With such a choice of this parameter we need only explicitly 
written terms in the relationship above in order to obtain the value of the 
integrand with linear accuracy in $\mu$. The same specification of this 
parameter also justifies consideration of only the two first terms in the 
expansion over $1/k^2$ above.

Calculating the low momentum integral we first expand function 
$\Phi^\mu_1(k)$ in the small parameter $\mu k$ (up to and including the 
term $(\mu k)^2\leq (\mu\sigma)^2$) 

\begin{equation}         \label{phi1exp}
\Phi^\mu_1(k)\approx \frac{1}{2}-\mu k+(\mu k)^2+O((\mu k)^3),
\end{equation}

and then perform the momentum integration

\begin{equation}     \label{low1}
{\widetilde F}^{<}(\mu) \approx 2\int_0^1 {dx}
\int_0^x {dy}  \frac{(1-x)^2}{x}  (1-2y) 
\biggl\{ ~\frac{1}{2}  \Bigl[~\ln{\frac{ \sigma^2 + a^2}{a^2}}
~-~ \frac{\sigma^2}{\sigma^2 + a^2}~\Bigr] 
\end{equation}
\[
-~\mu  \Bigl[~2 \sigma ~+~ \frac{\sigma  a^2}{\sigma^2 + a^2}
~-~ 3 a  \arctan{\frac{\sigma}{a}}~ \Bigr] ~
+~\mu^2  \Bigl[~ \sigma^2
~-~ \frac{\sigma^2 a^2}{\sigma^2 + a^2}
~- ~2 a^2 \ln{\frac{ \sigma^2 + a^2}{a^2}}
~\Bigl]~ \biggr\}.
\]

Calculating remaining integrals and omitting the higher order terms with the 
help of the condition $\sigma\simeq 1/\sqrt\mu$ we obtain

\begin{equation}                 \label{low2}
{\widetilde F}^{<} (\mu) ~\approx ~ 2  \Bigl[~-~ \frac{1}{48}
~+~ \frac{1}{4}  \ln{\sigma} ~+~
  \frac{15 \pi^2}{128}\mu ~
-~ \frac{\mu \sigma}{2} ~+~ \frac{(\mu \sigma )^2}{4} ~+~
\frac{1}{\sigma^2}  \Bigl(~ \frac{1}{6}  \ln{\sigma} ~+~
\frac{1}{12}~ \Bigr)~\Bigr].
\end{equation}

All terms with the auxiliary parameter $\sigma$ cancel in the sum of the 
high- and low-momentum integrals and we obtain with linear accuracy in 
the small parameter $\mu$

\begin{equation}  \label{tildemuterm}
{\widetilde F}(\mu) ~\approx ~ \Bigl[~-~\frac{1}{2}  \ln{(2 \mu)}
~-~ \frac{7}{24}~\Bigr] ~+~ \frac{15 \pi^2}{64}\mu.
\end{equation}

The first term in this relationship gives the contribution of the first 
order in the small mass ratio to hyperfine splitting, and is of no interest 
to us. The second term leads to the correction of order $(m/M)^2$ and 
calculation of such terms is the main goal of this work.

Let us consider further the origin of this term. First, 
it is not connected with the high-momentum integral, which generated only 
corrections of the previous order  in the mass ratio. The term which 
generates the correction under consideration in the low-momentum integral in 
Eq.\ (\ref{low1}) and Eq.\ (\ref{low2}) may be easily identified. This is 
the term which explicitly contains factor $\mu$ generated by the 
low-momentum expansion (see Eq.\ (\ref{phi1exp})) of the function 
$\Phi^\mu_1(k)$, and moreover it is the only term of order $\mu$ in the 
integral which remains finite when $\sigma$ goes to infinity. We may be even 
more concrete.  It is easy to see that the only term which remains finite 
when the upper limit in the integral $\int_0^\sigma dk^2 k^2/(k^2+a^2)^n \mu 
k$ goes to infinity has the characteristic form of $\arctan(\sigma/a)$. 

Now we are ready to describe a recipe for obtaining the term of order $\mu$ 
in the integral in Eq.\ (\ref{fmuint}) almost without calculations. First, 
we have to calculate only the low-momentum contribution to the integral in 
Eq.\ (\ref{fmuint}), and consider in the integrand only the term linear in 
$\mu$ generated by the respective term in the low-momentum expansion of 
the function $\Phi^\mu_1(k)$. At the second step we calculate the momentum 
integral and preserve in the result only contribution of the arctangent 
which gives factor $\pi/2$. At the last step we have to calculate 
the remaining double integral over the Feynman parameters. It is easy to 
check that this recipe immediately reproduces the term linear in $\mu$ 
in Eq.\ (\ref{tildemuterm}).

Up to now we have calculated only the linear in $\mu$ contribution to the 
integral ${\widetilde F}(\mu)$ in Eq.\ (\ref{fmuint}). However, it is easy 
to see that the recipe for extracting linear in $\mu$ contributions which we 
have just formulated is equally applicable for the total integral $F(\mu)$.  
The only difference between the integrals with the weight functions 
$bc_7$ and $\widetilde{bc_7}$ is in the details of the calculation of the 
high momentum contribution to the integral. The representation of the weight 
function in the form of the sum in Eq.\ (\ref{tilderepr}), simplified the 
calculation of the high-momentum contribution generated by the weight 
function $bc_7-\widetilde{bc_7}$, but as we have just seen the high-momentum 
contribution does not generate linear in $\mu$ contributions to the integral 
$F(\mu)$. Hence, we may apply the recipe  above for calculation of the 
total contribution of order $\mu$ to the integral $F(\mu)$, the only 
difference being that the role of the weight function will play now the 
total weight function $bc_7$.

It is clear that the same recipe will work each time when the $\mu$-integral 
for the contribution to the hyperfine splitting contains function 
$\Phi^\mu_1(k)$ in the integrand. Moreover, one may check explicitly that 
the recipe is valid even when the $\mu$-integral depends on other functions 
$\Phi^\mu_{0,2}(k)$, as happens for some contributions generated by 
the electron factor.  All respective calculations in the main body of the 
paper are made with the help of this recipe.

\section{Standard high-momentum integrals}
\label{appc}

As was mentioned in the main text and Appendix \ref{appb}
all high-momentum contributions to hyperfine splitting may written in terms  
of the standard  integrals 

\begin{equation}
V_{lmn} ~= ~ \int_{\sigma}^{\infty} \frac{d k^2}{(k^2)^l}
 (\ln k )^m  ~\Phi_n^{\mu} (k), 
\end{equation}

where $l=1,2$, $m=0,1$ and $n=1,2$.

These integrals should be calculated with linear accuracy in $\mu$. As was 
discussed in Appendix \ref{appb} this means that integrals should be 
calculated up to and including small terms of order $(\mu\sigma)^2$ (we 
remind the reader that $\sigma\approx1/\sqrt\mu$).

Consider in more detail the integrals with $n=1$

\begin{equation}
V_{lm1} ~=
~ \int_{\sigma}^{\infty} \frac{d k^2}{(k^2)^l}
 (\ln k )^m 
\Bigl[~-~ \mu  k   \sqrt {1 + \mu^2 k^2 } ~+~ \mu^2 k^2
~+~ \frac{1}{2} ~ \Bigr].
\end{equation}

First, we change the integration variable

\begin{equation}
t ~=~ \sqrt {1 + \mu^2 k^2 } ~-~ \mu k.
\end{equation}

In terms of this new variable typical integrals acquire the form

\begin{equation}
V_{101} ~=~ \int_0^{t_{\sigma}} {dt} ~\frac{t~(1 + t^2)}{1 - t^2},
\end{equation}

\begin{equation}
V_{111} ~=~ \int_0^{t_{\sigma}} {dt} ~\frac{t~(1 + t^2)}{1 - t^2}
 \ln { \frac{1 - t^2}{2 \mu t}} ,
\end{equation}

\begin{equation}
V_{201} ~=~ 4 \mu^2  \int_0^{t_{\sigma}} {dt}
~\frac{t^3~(1 + t^2)}{(1 - t^2)^3},
\end{equation}

\begin{equation}
V_{211} ~=~ 4 \mu^2 
\int_0^{t_{\sigma}} {dt} ~\frac{t^3~(1 + t^2)}{(1 - t^2)^3}
 \ln { \frac{1 - t^2}{2 \mu t}},
\end{equation}

where $t_{\sigma}=\sqrt {1 + \mu^2 \sigma^2 }- \mu \sigma\approx
1- \mu\sigma+ (\mu^2 \sigma^2)/{2}$.

Let us, for example, calculate the integral $V_{111}$

\begin{equation}  \label{v111}
V_{111} ~=~ \int_0^{t_0} {dt} ~t
\Bigl( ~\frac{2}{1 - t^2} ~-~ 1 ~ \Bigr)
 \ln { \frac{1 - t^2}{2 \mu t}} ~ \equiv ~
V^{'} ~+~ V^{''}.
\end{equation}

The first term in Eq.\ (\ref{v111}) may be written in the form

\begin{equation}
V{'}~=~ -~\int_0^{t_{\sigma}} {d \ln {(1 - t^2)}} 
\ln { \frac{1 - t^2}{2 \mu t}} 
\end{equation}
\[ 
=~-~\frac{1}{2}  \ln^2 {(1 - t_{\sigma}^2)} ~+~
\ln {( 2 \mu t_{\sigma})}  \ln {(1 - t_{\sigma}^2)}  ~-~
\int_0^{t_{\sigma}} \frac{d t}{t} \ln {(1 - t^2)}.
\]

With the help of an exact relationship 

\begin{equation}
1 - t_{\sigma}^2 ~=~ 2 \mu \sigma  t_{\sigma},
\end{equation}

$V'$ may be written in the form

\begin{equation}
V{'}~=~ \frac{1}{2}  \ln^2 {(2 \mu)} ~-~
\frac{1}{2}  \ln^2 {\sigma} ~+~
\frac{1}{2}  \ln^2 {t_{\sigma}} ~+~
\ln {(2 \mu)}   \ln {t_{\sigma}} 
\end{equation}
\[
-~\int_0^1 \frac{d t}{t}  \ln {(1 - t^2)} ~
+~\int_{t_{\sigma}}^1 \frac{d t}{t}  \ln {(1 - t^2)} .
\]

Expansion over  $(\mu\sigma)$ is facilitated by the relationship

\begin{equation}
\ln {t_{\sigma}} ~ \approx ~ -~ \mu \sigma
~+~ O[~(\mu \sigma )^3~],
\end{equation}

and we obtain

\begin{equation}
V^{'}~\approx~ \frac{1}{2}  \ln^2 {(2 \mu)} ~-~
\frac{1}{2}  \ln^2 {\sigma} ~+~
\frac{\pi^2}{12} ~+
~ (\mu \sigma)  \Bigl(~\ln{\sigma} ~-~ 1 ~\Bigr)
~+~ O[~(\mu \sigma )^3~] .
\end{equation}

Approximate expression for the integral $V^{''}$ may be obtained in the same 
way

\begin{equation}
V^{''} ~\approx ~ \frac{1}{4} ~+~ \frac{1}{2}  \ln {(2 \mu)} ~+~
(\mu \sigma)  \Bigl(~\ln{\sigma} ~-~ 1 ~\Bigr) ~+
~(\mu \sigma)^2  \Bigl(~-~\ln{\sigma} ~+~
\frac{1}{2} ~\Bigr),
\end{equation}

and then we obtain an approximation for the integral in Eq.\ (\ref{v111})

\begin{equation}
V_{111} ~\simeq ~ \frac{1}{2}  \ln^2 {(2 \mu)} ~+~
\frac{1}{2}  \ln {(2 \mu)} ~+~
\frac{\pi^2}{12} ~+~ \frac{1}{4}
~-~ \frac{1}{2}  \ln^2 {\sigma} 
\end{equation}
\[
+~(\mu \sigma)  \Bigl(~2 \ln{\sigma} ~-~ 2 ~\Bigr)
~+~ (\mu \sigma)^2  \Bigl(~-~ \ln{\sigma} ~+~ \frac{1}{2}
~\Bigr) .
\]

Approximate expressions for other integrals $V_{lmn}$ may be obtained in the 
same way, and they are collected in Table \ref{table3}.

\begin{table}
\caption{Coefficients in the Electron-Line Factor}
\begin{tabular}{ll}    
\\ 
$\mbox{c}_{1}$& $\frac{16}{y(1-y)^3}  
\Big[~(1-x)(x-3y)~-~2y\ln{x}~\Big]~$    
\\ \\ \tableline
\\
$\mbox{c}_{2}$      &   $\frac{4}{y(1-y)^3} 
\Big[~-(1-x)(x - y - 2y^2 /x)~+~2(x - 4y + 4y^2 /x)  \ln{x}~\Big]~$   
\\ \\ \tableline
\\
$\mbox{c}_{3}$     &   $\frac{1}{y(1-y)^2} 
\Big[~1 - 6x - 2x^2 -(y/x) (26 - 6y/x - 37x - 2x^2
+ 12xy + 16 \ln{x})~\Big]~$    
\\ \\ \tableline
\\
$\mbox{c}_{4}$   & $\frac{1}{y(1-y)^2} 
\Big(~2x - 4x^2 - 5y + 7xy~ \Big)~$     
\\ \\ \tableline
\\
$\mbox{c}_{5}$   & $\frac{1}{y(1-y)^2} 
\Big(~ 6x - 3x^2 - 8y + 2xy ~\Big)~$ 
\\ \\ \tableline
\\
$\mbox{c}_{6}$   & $-b^2  \frac{x - y}{x^2}~$    
\\ \\ \tableline
\\
$\mbox{c}_{7}$  & $2  \frac{1-x}{x}~$   
\\ \\
\end{tabular} 
\label{table1}
\end{table}

\narrowtext
\begin{table}
\caption{Electron-Factor Contributions to Hyperfine Splitting}
\begin{tabular}{lr}    \\
\ $\delta E_i$  & $\alpha(Z\alpha)\bigl(\frac{m}{M}\bigr)^2E_F$
\\ \\
\tableline   \\                                               
$\delta E_{123}$& $(-6 \ln2 - \frac{39}{32})~$    
\\ \\ \tableline
\\
$\delta E_{4}$   & $\frac{3}{16} ~$     
\\ \\ \tableline
\\
$\delta E_{5}$   & $-\frac{3}{32}~$ 
\\ \\ \tableline
\\
$\delta E_{6}$   & $-\frac{3}{32}~$    
\\ \\ \tableline
\\
$\delta E_{7}$  & $-\frac{9}{32}~$   
\\ \\
\tableline \tableline
\\
$\delta E_{\text {el}}=\sum_{i=1}^7\delta E_i$  & $(-6  \ln2 ~-~ 
\frac{3}{2}~)$ 
\\ 
\\ 
\end{tabular} 
\label{table2} 
\end{table}

\widetext
\begin{table}
\caption{Approximate expressions for the integrals $V_{lmn}$}
\begin{tabular}{ll}    
\\ 
$V_{100}$& $2\ln {(2 \mu)} ~-~ 2 ~+~ 2\ln {\sigma}~-\,(\mu \sigma)$    
\\ \\ \tableline
\\
$V_{110}$      &   $-~\ln^2 {(2 \mu)} ~+~ 2 \ln {(2 \mu)}~-~\frac{\pi^2}{6} 
~ -~ 2+~ \ln^2 {\sigma} ~+~(\mu \sigma)\Bigl(~- \ln{\sigma} ~+~ 1 ~\Bigr)$   
\\ \\ \tableline
\\
$V_{101}$     &   $-~\ln {(2 \mu)} ~-~ \frac{1}{2} ~-~
\ln {\sigma} ~+~ 2 (\mu \sigma) ~-~ (\mu \sigma)^2$    
\\ \\ \tableline
\\
$V_{111}$   & $\frac{1}{2}  \ln^2 {(2 \mu)} +
\frac{1}{2}  \ln {(2 \mu)} +
\frac{\pi^2}{12} + \frac{1}{4}
- \frac{1}{2}  \ln^2 {\sigma} +
 (\mu \sigma)  \Bigl(2 \ln{\sigma} - 2 \Bigr)
+ (\mu \sigma)^2  \Bigl(- \ln{\sigma} + \frac{1}{2}
\Bigr)$     
\\ \\ \tableline
\\
$V_{200}$   & $-\frac{1}{\sigma^2}$ 
\\ \\ \tableline
\\
$V_{201}$   & $\frac{1}{2\sigma^2}$    
\\ \\ \tableline
\\
$V_{202}$  & $\frac{1}{8\sigma^2}$   
\\ \\ \tableline
\\
$V_{210}$   & $-\frac{1}{\sigma^2} \cdot \Bigl(~\ln{\sigma} ~
+~ \frac{1}{2}~ \Bigr)$    
\\ \\ \tableline
\\
$V_{211}$  & $\frac{1}{2\sigma^2} \cdot \Bigl(~\ln{\sigma} ~
+~ \frac{1}{2}~ \Bigr)$   
\\ \\ \tableline
\\
$V_{212}$  & $\frac{1}{8\sigma^2} \cdot \Bigl(~\ln{\sigma} ~
+~ \frac{1}{2}~ \Bigr)$   
\\ \\
\end{tabular} 
\label{table3}
\end{table}

\end{document}